\pdfoutput=1
\documentclass[conference]{IEEEtran}
\IEEEoverridecommandlockouts
\usepackage[nocompress]{cite}
\usepackage{algpseudocode}
\usepackage{algorithm}
\usepackage{amsmath}
\usepackage[inline,shortlabels]{enumitem}
\usepackage[hidelinks]{hyperref}
\usepackage[capitalise]{cleveref}
\usepackage{pgfplots}
\usepackage{pgfplotstable}
\usepackage{tikz}
\usepackage[table]{xcolor} %
\usepackage{tcolorbox}
\usepackage{booktabs}
\usepackage{pifont}
\usepackage{subcaption}
\usepackage{microtype}
\usepackage{listings}
\usepackage{minted}
\tcbuselibrary{minted}
\usepackage{xcolor}
\definecolor{LightGray}{rgb}{0.95,0.95,0.95}

\usetikzlibrary{patterns}
\usepgfplotslibrary{groupplots}
\pgfplotsset{compat=1.18}


\usepackage{url}
\usepackage{balance}
\usepackage{microtype}
\usepackage{fancyhdr}

\usetikzlibrary{shapes.geometric, arrows, positioning}

\newcommand{\circl}[1]{%
  \tikz[baseline=(char.base)]{%
    \node[shape=circle,draw,inner sep=1pt] (char) {#1};}}

\include{plots/pgplots-definitions}

\begin{document}

\title{STORE: Self-Provisioning Storage-as-a-Service for Serverless Functions}


\author{
        \IEEEauthorblockN{Florian Trimmel}
        \IEEEauthorblockA{\textit{Distributed Systems Group} \\TU Wien, Vienna, Austria \\
            e12123438@student.tuwien.ac.at}
        \and
        \IEEEauthorblockN{Cynthia Marcelino}
        \IEEEauthorblockA{\textit{Distributed Systems Group} \\TU Wien, Vienna, Austria \\
            c.marcelino@dsg.tuwien.ac.at
        }
        \and
        \IEEEauthorblockN{Thomas Pusztai}
        \IEEEauthorblockA{\textit{Distributed Systems Group} \\TU Wien, Vienna, Austria \\
            t.pusztai@dsg.tuwien.ac.at
        }
        \and
        \IEEEauthorblockN{Stefan Nastic}
        \IEEEauthorblockA{\textit{Distributed Systems Group} \\TU Wien, Vienna, Austria \\
            snastic@dsg.tuwien.ac.at}
    }
\maketitle

\begin{abstract}
Serverless computing provides on-demand elasticity, pay-per-use, and simplified deployment. However, serverless functions are typically stateless and depend on external storage services such as object stores or databases to exchange data or support stateful functions. Provisioning and configuring these storage systems still requires manual setup or declarative scripts, introducing complexity, slowing development, and increasing the risk of configuration errors. 
To address these challenges, in this paper, we introduce \textsc{STORE}, a self-provisioning storage architecture for serverless functions. STORE automatically selects the optimal storage backend and eliminates developer effort through zero-touch and zero-configuration provisioning, achieved by moving the self-provisioning logic to the platform level. 
Our evaluation results show that STORE reduces implementation effort by up to 84\% compared to well-established Infrastructure-as-Code frameworks such as Terraform and Pulumi while maintaining low latency and linear scalability under realistic workloads, without introducing performance overhead.
\end{abstract}

\thispagestyle{plain}
\pagestyle{plain}



\section{Introduction}

In the serverless computing paradigm, the platform abstracts the infrastructure management, while the developer writes tiny pieces of code wrapped in small functions. Thus, serverless functions are typically stateless and rely on Backend-as-a-Services (BaaS) for additional features such as data management, which means they leverage external remote services such as Key-Value-Store (KVS), Object Store such as S3 and MinIO, and databases~\cite{scf,CloudProgrammingSimplified,goldfish2024,PerformanceIsolation,Lumos2025,Databelt2025}. Scaling out the application and supporting asynchronous, event-driven communication between components are inherent properties of serverless computing. Although serverless platforms abstract the infrastructure management and provide easy deployment, developers still face manual setup and configuration for essential components such as object storage and
databases, which are crucial for functions in a workflow to exchange data. The steps to provision a storage BaaS include provisioning, permissions, and integration, which add complexity and slow down the development process~\cite{ristov2024code,sokolowski2021automating,2024selfprovisioningInfrastructure}. 
The recently introduced concept of Self-Provisioning Infrastructure (SPI)~\cite{2024selfprovisioningInfrastructure} aims to fortify and extend the core strengths of serverless computing to also encompass BaaS, such as storage services, thereby eliminating provisioning efforts and reducing conceptual complexity to a pure focus on storage usage.

Common approaches that abstract BaaS provisioning include:
\begin{enumerate*}[label=(\alph*)]
\item \emph{Declarative Infrastructure} or Infrastructure as Code (IaC)~\cite{terraformTerraformHashiCorp,pulumiPulumiInfrastructure,awsCloudFormation} offers abstraction by specifying infrastructure in a declarative format, enabling automated provisioning and reproducible deployments. While IaC simplifies infrastructure management, it still requires developers to configure and maintain deployment scripts, which introduces complexity and potential configuration errors. Furthermore, IaC operates at provisioning time and does not impose runtime overhead on the deployed system. However, this static nature also limits adaptability: once applied, IaC definitions do not inherently support dynamic runtime adaptation to changing workloads or environmental conditions without external orchestration, re-deployment, or additional control mechanisms.

    \item\emph{Portability Models}~\cite{YussupovBKL20,WursterBFKLSS20} describe serverless applications in an agnostic pipes-and-filters architecture and use static code analysis to detect patterns such as storage usage. While these models address portability across providers, they still require infrastructure models to be explicitly defined and maintained.
    \item\emph{Serverless Databases}~\cite{DBLP:conf/icde/KesavanGTSM23, DBLP:conf/sigmod/SwensonKPTLSBBT25} are data sources that abstract database management as well as scalability and elasticity from the developer and provide a similar pay-per-use business model to FaaS offerings. While serverless databases lower the burden of infrastructure management, they are typically bound to a single underlying storage technology and do not typically abstract tasks such as user management and schema definitions.
\end{enumerate*}

While existing state-of-the-art provides a certain degree of abstraction, it nevertheless requires active developer involvement in configuring the environment. This introduces both qualitative and quantitative overheads in development time, as developers must identify and apply the appropriate system configurations. To address this limitation, we introduce STORE, a self-provisioning storage architecture for serverless functions that eliminates developer effort by enabling zero-touch and zero-configuration operation.  We summarize our contributions as follows:

\begin{itemize}
    \item \textbf{STORE:} A novel lifecycle model and architecture for self-provisioning serverless storage that defines lifecycle phases (preparation, resolution, policy enforcement, dynamic binding, translation). This enables automated provisioning and management of storage resources, eliminates manual configuration, and facilitates the serverless storage paradigm. Overall, STORE reduces implementation effort by up to 84\% with relative overheads as low as 0.19\%.
    \item \textbf{Dynamic Storage Selection:} A novel runtime mechanism for adaptive storage binding that automatically selects the optimal storage backend based on data shape, access pattern, and SLOs, ensuring efficient resource usage by matching workloads with the most suitable storage type without developer intervention. The mechanism also addresses the challenges of seamlessly redirecting or migrating workloads without service interruption and can additionally optimize for target metrics based on provider scoring.
    \item \textbf{Zero-Touch Configuration:} A developer-transparent provisioning mechanism that removes the need for manual setup. STORE achieves this by embedding self-provisioning logic at the serverless platform level, enabling functions to seamlessly access storage while enforcing fine-grained permissions, ensuring zero-touch, zero-configuration operation.
\end{itemize}

This paper has six sections. 
\cref{related_work} presents related work. 
\cref{architecture} describes the Store lifecycle as well as the architecture overview. 
\cref{mechanisms} describes the dynamic storage selection and the zero-touch configuration introduced by STORE and their usage. 
\cref{evaluation} discusses the experiments and evaluation, 
\cref{conclusion} concludes with a final discussion and future work.

\section{Related Work}\label{related_work}


\subsection{Declarative Infrastructure}

First-~\cite{awsCloudFormation, terraformTerraformHashiCorp} and second-generation~\cite{amazonCloudDevelopment, pulumiPulumiInfrastructure} infrastructure-as-code tools are already well-established and widely used. Those, however, have the shortcomings of still requiring manual specification of the desired system state, as well as a limited ability to react to dynamic system changes and complex requirements. 
The concept of intent-based infrastructure~\cite{Allam_2025,DBLP:conf/nof/BezahafHBDBKH19,DBLP:journals/sensors/AndradeHozWA24} aims to address these shortcomings by enabling the specification of higher-level user intents in the system, which are evaluated and acted upon based on telemetry to achieve a state that matches the user's intent. While intent-based infrastructure reduces developer effort, it still requires explicit specification of intents, which are then extended and translated into lower-level configuration, often via AI~\cite{DBLP:conf/nof/BezahafHBDBKH19,DBLP:journals/sensors/AndradeHozWA24}.
While approaches such as CloudCAMP~\cite{DBLP:conf/ucc/BhattacharjeeBG18} and ARGON~\cite{DBLP:conf/models/SandobalinIA19} promise end-to-end provisioning, they rely on defining either a system architecture (to be converted into infrastructure definitions) or the required infrastructure itself. Additionally, most model-based approaches operate on a higher level of abstraction and produce IaC as output, still requiring interaction with IaC tools.
Darklang~\cite{darklangDarklang} and Wing~\cite{winglangWingProgramming} promise to completely eliminate infrastructure declaration. Darklang acts as a cloud provider. Thus, while no provisioning is necessary, a type of vendor lock-in to Darklang cloud infrastructure still exists. Wing is a cloud-native programming language that intermingles infrastructure and business logic code. While tightly integrated, the declaration of infrastructure is still necessary, and the translation step from Wing code to lower-level IaC provides an interface for IaC tools.

Although declarative and intent-based approaches reduce effort, they still rely on explicit infrastructure specifications or new programming models. STORE eliminates infrastructure declarations entirely, abstracting storage at the level of workflows and data shapes, and letting the platform handle provisioning and adaptation transparently.

\subsection{Portability Models}

Function-centric portability models~\cite{DBLP:conf/seke/QianZ20} allow reducing function code to a minimum, providing portability and composability across different cloud providers. Those frameworks tend to restrict interaction with BaaS to provide simplified views on functions and, therefore, do not usually interact with storage.
Cloud Service Abstraction, promising portability by introducing an abstraction layer above BaaS, has been explored for serverless use cases~\cite{DBLP:conf/cloudcom/MoCCL23}. While abstraction libraries such as Libcloud~\cite{apacheApacheLibcloud} or jclouds~\cite{apacheApacheJcloudsxAE} address vendor lock-in, the actual provisioning of infrastructure still has to be done manually.
Model-based approaches provision federated serverless environments by abstracting primitives across providers but require extensive models and often depend on IaC for execution. For example, UMLPMSC~\cite{DBLP:conf/icsca/SameaAAKR19} defines stereotypes for storage and databases that map to provider resources.
 \textsc{BaaSLess}~\cite{LarcherGNR24} enables the dynamic provisioning of storage buckets from two different storage providers, focusing on federated systems where the supporting BaaS may be constrained by provider-specific requirements, whereas STORE focuses on more general aspects of provisioning. 
 Similarly, \textsc{StoreLess}~\cite{DBLP:conf/icsoc/RistovHGNPF24} focuses on selecting the optimal storage type in a federated serverless environment by leveraging colocation and a list-based heuristic. \textsc{StoreLess} focuses on selecting the best storage region, whereas our solution focuses on selecting the best storage type.

Although portability models aim to avoid vendor lock-in, they typically require extensive upfront modeling and static analysis, which require predefined abstractions that must be maintained over time. In contrast, STORE dynamically provisions and selects storage at runtime, adapting to workload characteristics without additional developer modeling effort.

\subsection{Serverless Storage}
Cloud provider offerings, such as Firestore~\cite{DBLP:conf/icde/KesavanGTSM23} and DynamoDB~\cite{280754}, face the issue of vendor lock-in to their respective platforms. FaunaDB, based on Calvin~\cite{DBLP:conf/sigmod/ThomsonDWRSA12}, is a deterministic database that provides serverless capabilities via an API. CockroachDB~Serverless~\cite{DBLP:conf/sigmod/SwensonKPTLSBBT25}, based on the open-source database CockroachDB, provides relational database capabilities via an SQL interface. All solutions abstract many database management tasks, particularly those related to elasticity and scalability. However, the last-mile effort, including configuring connections, setting permissions, defining schemas, learning the APIs, and formulating SQL queries, still remains.

While serverless databases and STORE share some similarities in terms of pay-as-you-go pricing and compute scaling, STORE is not an implementation of a serverless database itself. STORE acts as an enabler, further simplifying serverless storage and combating vendor lock-in while also providing zero-touch configuration. Serverless databases can be easily integrated into STORE as an underlying storage via a pluggable translation component.

\section{STORE Lifecycle Phases and Architecture Overview}\label{architecture}
STORE introduces self-provisioning storage, shifting BaaS management from developers to the platform. Provisioning, scaling, and placement are fully abstracted away, allowing the system to autonomously allocate and manage data across the continuum based on access patterns, consistency requirements, and cost–performance trade-offs. Persisting or retrieving data should require no more than a single, intuitive API call, independent of when or where the function executes.
By unifying compute and storage semantics for serverless functions, STORE enables developers to focus solely on what their functions do, while the platform transparently manages how and where both computation and data reside.
To realize our vision of self-provisioning storage, we present lifecycle phases that enable it, along with an overview of the architecture that implements them. 


\subsection{Self-Provisioning Lifecycle Phases} 
\begin{figure}[t]
    \centering
    \includegraphics[width=\linewidth]{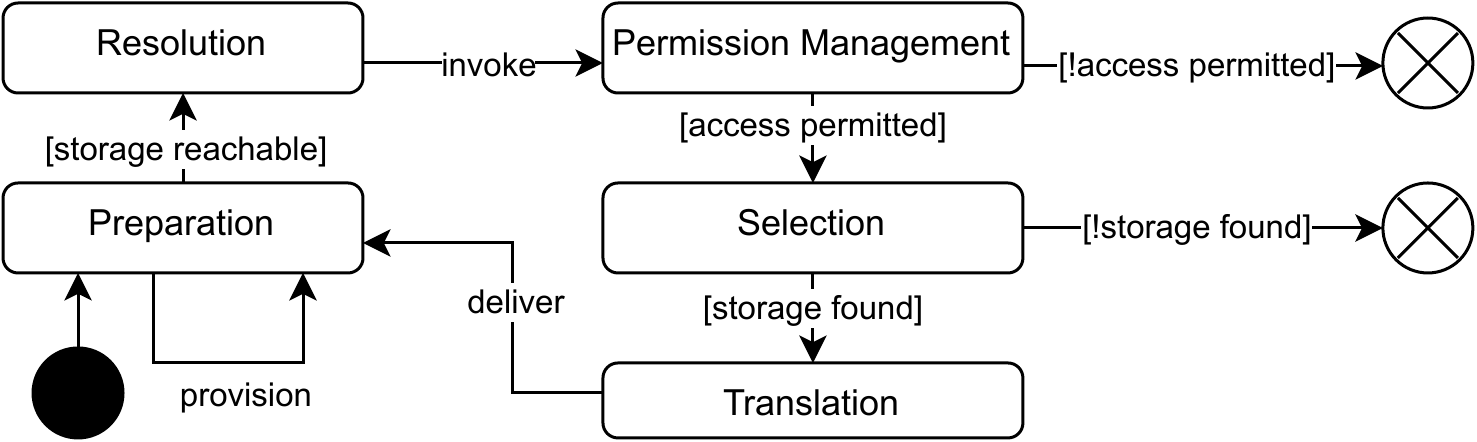}
    \caption{Self Provisioning Lifecycle Phases}
    \label{fig:phases}
\end{figure}

As shown in \cref{fig:phases}, \textsc{STORE} self-provisioning happens in five phases, from preparation to translation. These phases are designed to automate provisioning, enforce security, and dynamically adapt storage. Each phase is responsible for specific tasks, as described below.

\paragraph{PREPARATION} In this phase, the groundwork for Self-Provisioning is laid out. The aim is to handle all tasks that can be done statically at deployment time. For example, it is essential to ensure that storage systems (e.g, MinIO) are in place and that communication with them is possible.

\paragraph{RESOLUTION} This phase is a part of enabling Zero-Touch/Zero-Configuration by ensuring that functions can communicate with a STORE implementation without the developer needing to provide connection details or authentication credentials. This can, in part, be done statically by instrumenting resources. Data that changes for every function call can only be resolved dynamically. 

\paragraph{PERMISSION MANAGEMENT} Two modes of permission management can be distinguished. One type, handled directly by a STORE implementation, enables uniform access checking across all storage. The second type, relayed to the underlying storage, ensures that other systems accessing it directly experience a similar level of security to when using it through our solution. Permissions are managed by security level, ensuring that only specific functions or workflows have access to the data.  

\paragraph{SELECTION} Based on selection criteria such as the size of data to store, usage statistics, or based on the structure of the data (for example, distinguishing between tabular and binary data), a suitable type of underlying storage is selected. Switching storage when data changes is also possible. 

\paragraph{TRANSLATION} In this phase, storage has already been selected. A received request has to be translated into (potentially multiple) new instructions for storing or retrieving data from the underlying storage. Translation can happen directly, as in storing tabular data in a column store, where minimal extra translation effort is required. Storing data with a structure that is not the primary objective of a storage system is possible, but it requires additional effort. For example, storing chunks of binary data in a column store requires careful serialization and creation of suitably typed columns.  

\subsection{STORE Architecture Overview}

\begin{figure}[t]
    \centering
    \includegraphics[width=\linewidth]{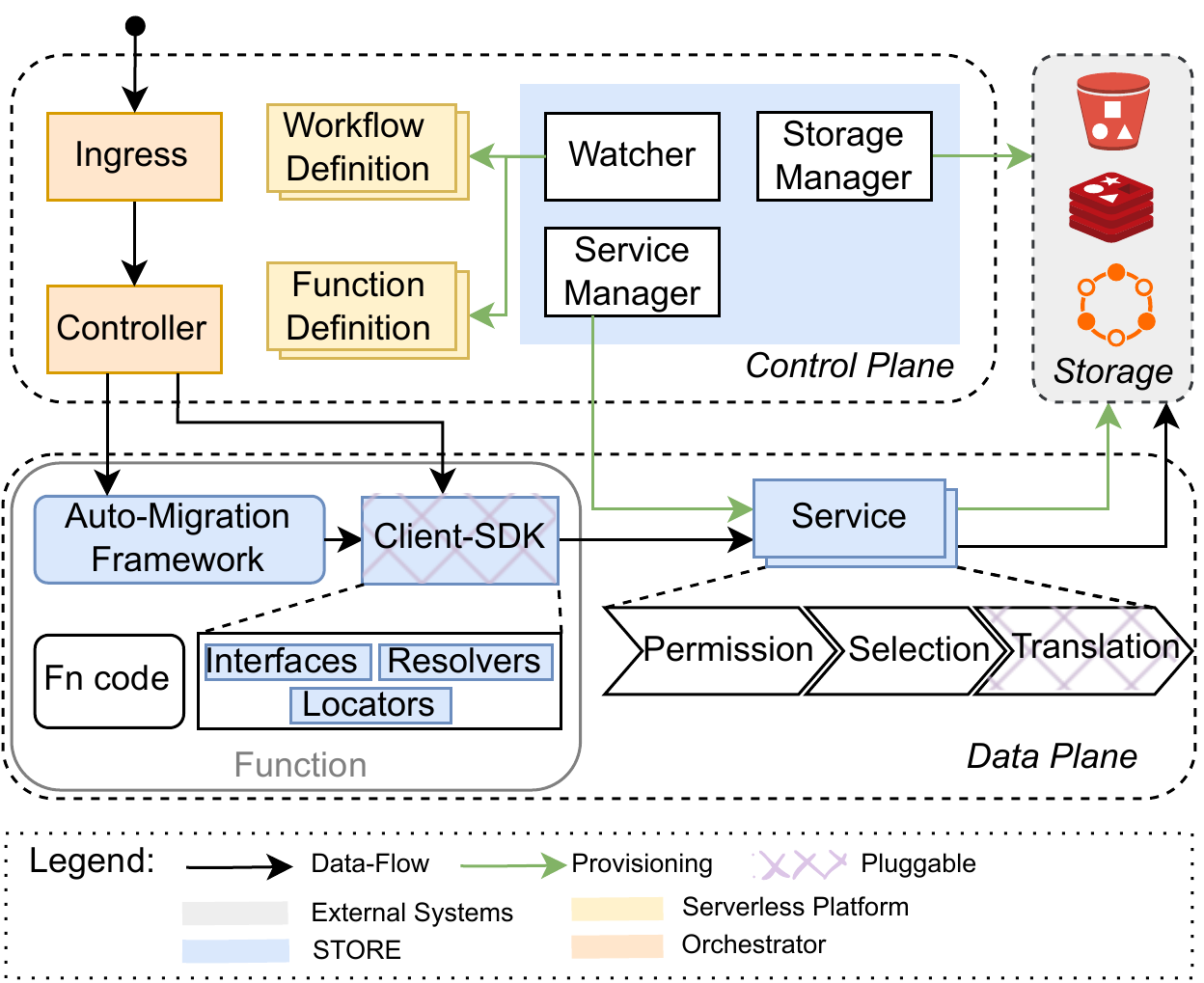}
    \caption{STORE Architecture Overview}
    \label{fig:store_overview}
\end{figure}

\cref{fig:store_overview} shows the overall architecture. STORE is designed as a set of modules on top of an orchestrator and integrated with a serverless platform. Pluggable components are introduced to facilitate integration with different serverless platforms and storage technologies. 

The control plane ensures the system is operated in the correct state. This includes ensuring communication pathways exist from the client to the storage, that the underlying storage is provisioned correctly, and that requests are routed to the called functions. It comprises a Watcher monitoring serverless resources, a STORE-Service Manager provisioning the STORE-Service, and a Storage Manager provisioning underlying storage.

The data plane, consisting of Client-SDK, Auto-Migration Framework, and STORE-Service, manages dynamic storage selection and is crucial for enabling zero-touch configuration.

\subsubsection{Components} Our architecture comprises six components: Watcher, Storage Manager, STORE-Service Manager, STORE-Service, Client-SDK, and Auto-Migration Framework.

\paragraph{Watcher} It connects to the underlying orchestrator and interacts with resources provided by the serverless platform. It aims to instrument serverless resources, such as functions and workflows, with the information required for Zero Touch Configuration. Configuration information is injected into serverless resources so it can be picked up by the Client-SDK. Events from the orchestrator, changes to the STORE-Service configuration, and the addition or modification of serverless resources trigger the reinjection of configuration information.

\paragraph{Storage Manager} It performs provisioning of underlying storage systems. It is integrated with the orchestrator through the definition of resources that depict storage. Those resources serve both as a definition of the desired state of the underlying storage system and collectively as the pool of available storage for the STORE-Service.

\paragraph{STORE-Service Manager} It manages configuration and performs provisioning of the STORE-Service instances. Receives events about configured storage and provides it to the STORE-Service. It is integrated with the orchestrator through resource definitions. An instance of the STORE-Service is created for each tenant of the system and linked to the available storage systems for this tenant.

\paragraph{STORE-Service} It lies at the heart of the system. It is responsible for the lifecycle phases of permission management, dynamic storage selection, and translating requests into the APIs of concrete storage systems. It is also responsible for validating the state of the underlying storage systems and for provisioning resources on them. Access control is enforced by the STORE-Service based on the identity information provided with each request. The Translation part of the STORE-Service is designed as a pluggable set of components, allowing for quick addition of additional types of underlying systems.

\paragraph{Client-SDK} It enables Zero-Touch integration when developing serverless functions. Retrieves data about the current system state, the location of the STORE-Service, and data about the context in which the function is run. The Client-SDK is designed as a high-level interface and a set of pluggable bindings that integrate with different development environments. The interfaces subcomponent encompasses various ways of interacting with the SDK, such as via a client or a map-like interface. Resolvers provide information about the function in the context of the system, for example, the function or workflow names, while locators allow the client to locate other system components, such as the STORE-Service. Resilience is supported through retries in the client, while function-level resilience is provided natively by the serverless platform.

\paragraph{Auto-Migration Framework} It allows for seamless migration from a wide variety of existing storage code to STORE by providing interface mimics for storage-specific libraries used in existing functions. Integrates with the Client-SDK by mapping used function calls to corresponding Client-SDK calls. A comprehensive set of Auto-Migration Libraries comprises the Auto-Migration Framework that enables backward compatibility.

\section{STORE Mechanisms}\label{mechanisms}

\subsection{Dynamic Storage Selection}

\begin{algorithm}[!b]
\caption{Dynamic Storage Selection}
\label{alg:dynamic-storage}
\begin{algorithmic}[1]
    \Require data.size $\neq \emptyset$, data.stream $\neq \emptyset$, data.structure     
    \Require providers (type, canStream, structure, stats) 
    \Require \textit{optional } typePreselect 
    \If{data.structure = $\emptyset$} \label{alg:dynamic-storage:1}
        \State data.structure $\gets$ \texttt{blob} \label{alg:dynamic-storage:2}
    \EndIf

    \If{typePreselect $\neq \emptyset$} \label{alg:dynamic-storage:4}
        \State providers $\gets$ filterProvidersByType() \label{alg:dynamic-storage:5}
    \EndIf

    \If{data.size $>$ STREAMING\_THRESHOLD} \label{alg:dynamic-storage:7}
        \State providers $\gets$ filterStreamingProviders() \label{alg:dynamic-storage:8}
    \EndIf

    \ForAll{$p \in$ providers} \label{alg:dynamic-storage:10}
        \If{$p.structure \neq data.structure$} \label{alg:dynamic-storage:11}
            \State $p.score \gets p.stats $-$ \text{TRANSLATION\_PENALTY}$ \label{alg:dynamic-storage:12}
        \Else \label{alg:dynamic-storage:13}
            \State $p.score \gets p.stats$ \label{alg:dynamic-storage:14}
        \EndIf
    \EndFor \label{alg:dynamic-storage:15}

    \State bestProvider $\gets$ argmax(providers, $p \to p.score$)   \label{alg:dynamic-storage:17}

    \State \Return bestProvider \label{alg:dynamic-storage:18}
\end{algorithmic}
\end{algorithm}

Dynamic Provider Selection aims to choose the best storage provider implementation based on different criteria. The structure of the data (e.g, tabular, binary), the data size, and usage statistics are taken into account. Data structure is used as a filter to discard incompatible storage providers. Data size is used to restrict providers that handle bigger data sizes suboptimally. Usage statistics are consolidated into a single, comparable metric and used to select the best-performing storage provider from the remaining options. \cref{alg:dynamic-storage} shows how the decision which provider to use is made.
The dynamic provider selection algorithm (\cref{alg:dynamic-storage}) receives as input the data size, data shape, and data structure. If the data shape is unknown, STORE selects the default storage selection, blob storage (\cref{alg:dynamic-storage:2}). Additionally, STORE provides the developer with the flexibility to preselect file-type storage providers (\cref{alg:dynamic-storage:4}).
The algorithm runs on the \textit{STORE-Service} and filters the list of available providers (\cref{alg:dynamic-storage:5,alg:dynamic-storage:8}). Due to the negative effects on latency, when storing large chunks of data synchronously, providers that do not support request streaming are excluded above a certain streaming threshold (\cref{alg:dynamic-storage:7}). Next, a translation penalty is applied for providers that do not natively support the data structure to be stored (\cref{alg:dynamic-storage:12}). Finally, the best-scoring provider among the remaining candidates is selected (\cref{alg:dynamic-storage:17}). A provider score can be adjusted based on sampling latency, cost, or the current fulfillment status of an SLO. 
Switching providers can be handled in two different ways. Either a switch penalty is introduced to account for additional provisioning effort, or a lazy approach is taken, where provisioning of the underlying storage begins only after the decision to switch is made. When the provisioning is finished, the next request triggers the switch. This way, no latency increase is incurred for provisioning the new storage destination.

\subsection{Zero-Touch Configuration}

\begin{figure}[!b]
    \centering
    \includegraphics[width=\linewidth]{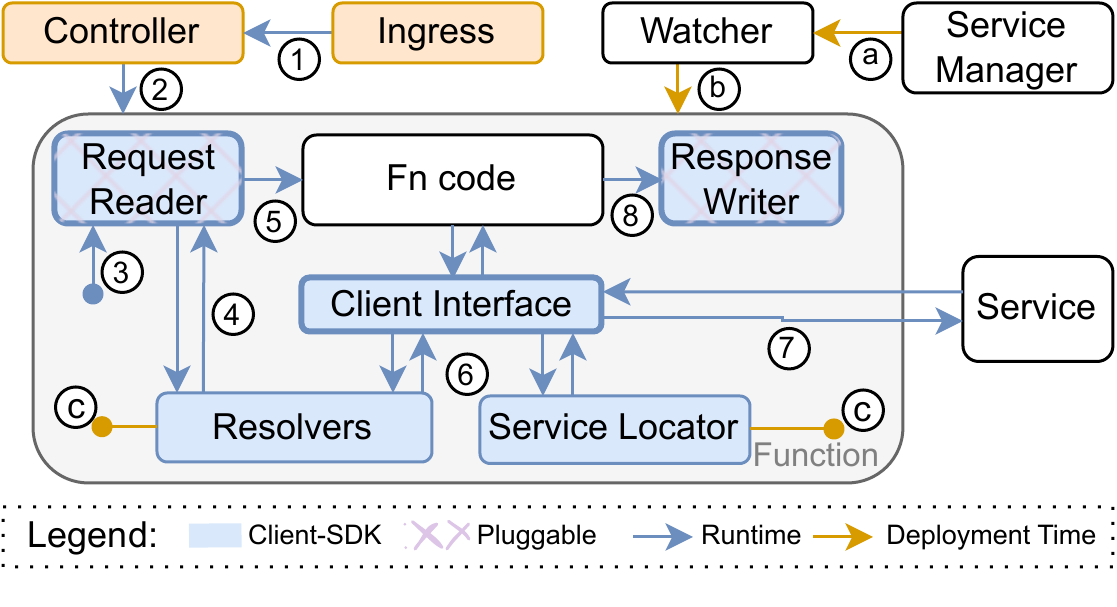}
    \caption{Zero-Touch Configuration Overview}
    \label{fig:zero-touch}
\end{figure}

The Zero-Touch Configuration mechanism is designed to eliminate configuration interaction between the developer and STORE. In our architecture, zero-touch configuration comprises deployment-time instrumentation, runtime resolution, and runtime data passing. \cref{fig:zero-touch} illustrates the key steps comprising Zero-Touch Configuration, split between deployment time and runtime. \circl{a} At deployment time, the STORE-Service Manager informs the watcher about the currently active STORE-Service instances and how they can be reached. \circl{b} The watcher then instruments the function and workflow definitions with information about the location of the STORE-Service and the identities of the functions. This includes the function name and workflow name, if present. \circl{c} The information is made available to the function instances in a way that the identity resolver and service locator components can retrieve it via runtime resolution. \circl{1} At runtime, a request arrives at the ingress (either as a function execution request or as an event for workflow execution) and is sent to the controller component, which provisions a function instance (in case of a cold start). \circl{c} The identity resolver and service locator parse their respective information after startup. \circl{2} The controller then relays the request to a function instance. \circl{3} The request reader picks up the request and starts runtime data processing, such as extracting the workflow execution ID. \circl{4} The information is then stored in the identity resolver for later pickup. Here, the information scope is relevant, and isolation per request must be provided. \circl{5} After extracting the necessary information, the request reader calls the function handler. In the function handler, the Client-SDK can be used without any prior configuration. \circl{6} The Client-SDK internally retrieves configuration information from the resolvers and the service locator. \circl{7} Service communication happens through the Client-SDK, storing and retrieving data as instructed by the function and passing identity information for access control. \circl{8} After the function returns a response body, the response writer inspects the returned value and enriches it with additional runtime-specific information, finishing runtime data passing.

\section{Evaluation}\label{evaluation}
The goal of the experiments is to assess whether our architecture alleviates developers of the burden of manually provisioning infrastructure without compromising the system's performance and scalability. The performance experiment varies the input size, while the scalability experiments vary the requests per second to assess how our system behaves under high load. These experiments present real-world invocation patterns as described in~\cite{CloudProgrammingSimplified}. Each experiment is repeated five times to reduce bias, and the reported results are calculated as the arithmetic mean across all repetitions. While STORE allows attaching arbitrary storage systems via the \textit{Storage Manager} and the pluggable \textit{Translation} components, the prototype implementation currently supports only binary and tabular data.

\paragraph{Experimental Workflows} We design multiple use cases to demonstrate our architecture’s core capabilities in realistic Serverless scenarios. They cover file and tabular storage, custom and workflow-scoped permissions, dynamic storage selection through the \texttt{map} interface, and concurrency control, showcasing how zero-touch configuration supports diverse serverless applications. 
\begin{enumerate*}
    \item \emph{Storing and Retrieving Files} Functions use the \texttt{Client-SDK} DSL interface, which enables explicit metadata configuration (e.g., security level \texttt{PUBLIC}), while both variants store or retrieve files by name and return either the result or an error message.
    \item \emph{Storing and Retrieving Tabular Data.}
    The storage function specifies fine-grained permissions by adding identities and allowing other functions to access data, demonstrating zero-touch identity resolution. Retrieval supports row- and column-level queries, ranging from single attributes to complete rows.
    \item \emph{Sequential Workflow} chains file and tabular operations in three steps, highlighting workflow- and execution-scoped permissions. 
\end{enumerate*}

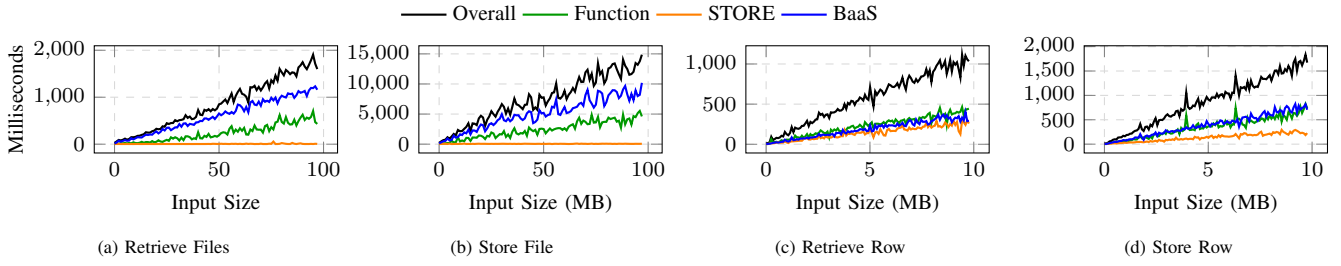
\begin{figure*}[t]
    \begin{subfigure}{0.24\linewidth}
      \begin{tikzpicture}
        \begin{axis}[
          xlabel={Input Size}, 
          ylabel={Milliseconds},
          ylabel style={yshift=-3pt,font=\footnotesize},
          xticklabel style={font=\footnotesize},
          xlabel style={font=\footnotesize},
          yticklabel style={font=\footnotesize},
          width=4.8cm, height=3cm,
          grid=major, grid style={dashed,gray!30},
          mark options={solid},
          legend style={at={(2.25,1.1)},anchor=south,draw=none,legend columns=-1, font=\footnotesize},
        ]
    
           \addplot+[no marks, color=black, thick]
          table[x=X, y=overall, col sep=comma]
          {experiments/performance/read_file.csv};
          \addplot+[no marks, color=green!60!black, thick] table[x=X, y={retrieve-function}, col sep=comma] {experiments/performance/read_file.csv};
          \addplot+[no marks, color=orange, thick] table[x=X, y={store-service},    col sep=comma] {experiments/performance/read_file.csv};
          \addplot+[no marks, color=blue, thick] table[x=X, y={minio.download},   col sep=comma] {experiments/performance/read_file.csv};
          \legend{Overall, Function, STORE, BaaS}
        \end{axis}
      \end{tikzpicture}
      \caption{Retrieve Files}
      \label{fig:read-files-performance}
    \end{subfigure}
       \begin{subfigure}{0.24\linewidth}
        \begin{tikzpicture}
            \begin{axis}[               
                xlabel={Input Size (MB)},
                ylabel style={yshift=-3pt,font=\footnotesize},
                ylabel={},
                xticklabel style={font=\footnotesize},  
                xlabel style={font=\footnotesize},
                yticklabel style={font=\footnotesize}, 
                width=4.8cm,                     
                height=3cm,
                grid=major,
                scaled y ticks=false,
                grid style={dashed,gray!30},
                mark options={solid}
            ]
           
            \addplot+[no marks, color=black, thick]
              table[x=X, y=overall, col sep=comma]
              {experiments/performance/store_file.csv};
              \addplot+[no marks, color=green!60!black, thick] table[x=X, y={store-function}, col sep=comma] {experiments/performance/store_file.csv};
              \addplot+[no marks, color=orange, thick] table[x=X, y={store-service},    col sep=comma] {experiments/performance/store_file.csv};
              \addplot+[no marks, color=blue, thick] table[x=X, y={minio.upload},   col sep=comma] {experiments/performance/store_file.csv};
            
            \end{axis}
        \end{tikzpicture}
        \caption{Store File}
        \label{fig:store-files-performance}
    \end{subfigure}
    \begin{subfigure}{0.24\linewidth}
        \begin{tikzpicture}
            \begin{axis}[               
                xlabel={Input Size (MB)},
                ylabel style={yshift=-3pt,font=\footnotesize},
                xticklabel style={font=\footnotesize},  
                xlabel style={font=\footnotesize},
                yticklabel style={font=\footnotesize}, 
                ylabel={},
                width=4.8cm,                    
                height=3cm,
                grid=major,
                grid style={dashed,gray!30},
                mark options={solid}
            ]
            \addplot+[no marks, color=black, thick]
          table[x=X, y=overall, col sep=comma]
          {experiments/performance/read_row.csv};
          \addplot+[no marks, color=green!60!black, thick] table[x=X, y={retrieve-row-function}, col sep=comma] {experiments/performance/read_row.csv};
          \addplot+[no marks, color=orange, thick] table[x=X, y={store-service},    col sep=comma] {experiments/performance/read_row.csv};
          \addplot+[no marks, color=blue, thick] table[x=X, y={cassandra.get},   col sep=comma] {experiments/performance/read_row.csv};
            \end{axis}
        \end{tikzpicture}
        \caption{Retrieve Row}
        \label{fig:read-rows-performance}
    \end{subfigure}
    \begin{subfigure}{0.24\linewidth}
        \begin{tikzpicture}
            \begin{axis}[               
                xlabel={Input Size (MB)},
                ylabel style={yshift=-6pt,font=\footnotesize},
                ylabel={},
                xticklabel style={font=\footnotesize},  
                xlabel style={font=\footnotesize},
                yticklabel style={font=\footnotesize}, 
                width=4.8cm,                      
                height=3cm,
                grid=major,
                grid style={dashed,gray!30},
                mark options={solid}
            ]
             \addplot+[no marks, color=black, thick]
          table[x=X, y=overall, col sep=comma]
          {experiments/performance/store_row.csv};
          \addplot+[no marks, color=green!60!black, thick] table[x=X, y={store-row-function}, col sep=comma] {experiments/performance/store_row.csv};
          \addplot+[no marks, color=orange, thick] table[x=X, y={store-service},    col sep=comma] {experiments/performance/store_row.csv};
          \addplot+[no marks, color=blue, thick] table[x=X, y={cassandra.save},   col sep=comma] {experiments/performance/store_row.csv};
            \end{axis}
        \end{tikzpicture}
        \caption{Store Row}
        \label{fig:store-rows-performance}
    \end{subfigure}
    \caption{Performance of serverless workflows with varying input sizes (MB) }
    \label{fig:performance}
\end{figure*}

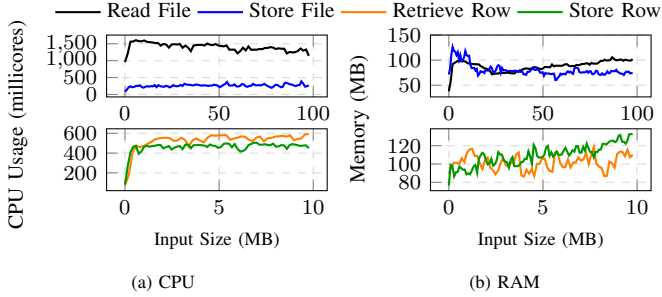
\begin{figure}[t]
    \begin{subfigure}{0.49\linewidth}
        \begin{tikzpicture}
            \begin{groupplot}[
              group style={group size=1 by 2, vertical sep=12pt},
              width=4.5cm, height=2.4cm,
              axis lines=box, ytick pos=left,
              yticklabel style={font=\footnotesize},
              xticklabel style={font=\footnotesize},  
              grid=major, grid style={dashed,gray!30},
              ylabel={CPU Usage (millicores)},
            ]
            
            \nextgroupplot[
              ylabel style={font=\footnotesize, xshift=-22pt},
              legend style={at={(1.15,1.7)}, anchor=north, draw=none, font=\footnotesize, legend columns=-1},
            ]
            \addplot+[no marks, color=black, thick]
              table[x=X, y={store}, col sep=comma]{experiments/performance/res_usage/data/read_file_store_cpu.csv};
            \addlegendentry{Read File}
            
            \addplot+[no marks, color=blue, thick]
              table[x=X, y={store}, col sep=comma]{experiments/performance/res_usage/data/store_file_store_cpu.csv};
            \addlegendentry{Store File}

            \addlegendimage{no markers, draw=orange, thick}
            \addlegendentry{Retrieve Row}

            \addlegendimage{no markers, draw=green!60!black, thick}
            \addlegendentry{Store Row}
            
            \nextgroupplot[axis lines=box, ytick pos=left, ylabel={},
            xlabel={Input Size (MB)},
            xlabel style={font=\scriptsize}]
            \addplot+[no marks, color=orange, thick]
              table[x=X, y={store}, col sep=comma]{experiments/performance/res_usage/data/read_row_store_cpu.csv};
            
            \addplot+[no marks, color=green!60!black, thick]
              table[x=X, y={store}, col sep=comma]{experiments/performance/res_usage/data/store_row_store_cpu.csv};
            
            \end{groupplot}
            
            \end{tikzpicture}
        \caption{CPU}
    \end{subfigure}
    \hfill
    \begin{subfigure}{0.49\linewidth}
        \begin{tikzpicture}
            \begin{groupplot}[
              group style={group size=1 by 2, vertical sep=12pt},
              width=4.5cm, height=2.4cm,
              axis lines=box, ytick pos=left,
              yticklabel style={font=\footnotesize},
              xticklabel style={font=\footnotesize},  
              grid=major, grid style={dashed,gray!30},
              ylabel={Memory (MB)}
            ]
            
            \nextgroupplot[
              ymax=150,
              ylabel style={font=\footnotesize, xshift=-20pt},
            ]
            \addplot+[no marks, color=black, thick]
              table[x=X, y={store}, col sep=comma]{experiments/performance/res_usage/data/read_file_store_mem.csv};
            
            \addplot+[no marks, color=blue, thick]
              table[x=X, y={store}, col sep=comma]{experiments/performance/res_usage/data/store_file_store_mem.csv};
            
            \nextgroupplot[axis lines=box, ytick pos=left, ylabel={},
            xlabel={Input Size (MB)},
            xlabel style={font=\scriptsize}
            ]
            \addplot+[no marks, color=orange, thick]
              table[x=X, y={store}, col sep=comma]{experiments/performance/res_usage/data/read_row_store_mem.csv};

            \addplot+[no marks, color=green!60!black, thick]
              table[x=X, y={store}, col sep=comma]{experiments/performance/res_usage/data/store_row_store_mem.csv};
            
            \end{groupplot}
            
            \end{tikzpicture}
        \caption{RAM}
    \end{subfigure}
    \caption{Resource usage of Self Provisioning Service where: (a) CPU, and (b) RAM. Solid line is Self Provisioning Service}
    \label{fig:performance-resources}
\end{figure}

\subsection{Experimental Setup}

The prototype is available as an open-source project on Github~\footnote{\url{https://github.com/polaris-slo-cloud/store}}. 
The \textsc{STORE} prototype is implemented in Kotlin.
Our experiments are executed on a cluster provisioned on a server with 32 CPUs, 378 GiB RAM, and SSD-backed ZFS storage. The cluster consists of seven Debian 12 VMs (one control plane, three dedicated storage nodes, and three worker nodes), running Kubernetes. Knative Serving and Eventing are deployed with autoscaling tuned to support high concurrency, while cold starts are disabled to avoid scale-to-zero penalties. MinIO~\cite{minio} and Cassandra~\cite{apacheCassandra} are installed on the storage nodes, configured for erasure coding and rack-level replication. Redis~\cite{redis} is used for persistent metadata and permissions storage. Observability is provided by Jaeger~\cite{jaeger} for distributed tracing and Prometheus~\cite{prometheus} with Grafana~\cite{grafana} for fine-grained CPU and memory monitoring at a 5s interval.


\subsection{Performance Results}

\subsubsection{Latency}


Overall, the experiments show that \textsc{STORE} introduces minor and predictable overhead for file-based and tabular workloads, validating its scalability across different storage types.

File-based workloads: For both storing and retrieving files, \textsc{STORE} contributes only a minor fraction of the total function's execution time from  0.19\% to 26.08\% (at 1~KB; rapidly decreasing)
\cref{fig:read-files-performance,fig:store-files-performance} show that the latency increases with input size due to the storage-bound nature of file operations, but the service overhead remains small and stable. This behavior results from the service's streaming design, which allows processing to begin without waiting for the full payload.

Tabular workloads: For storing and retrieving rows, the \textsc{STORE}-Service contributes to the total latency from 12.56\% to 31.79\%  
\cref{fig:read-rows-performance,fig:store-rows-performance} illustrate that the service (in contrast to file-based workloads) must fully read and translate the request payload into CQL statements. The latency difference can be attributed to the implemented \textit{Translation} component not supporting request streaming. Despite this, overall latency scales linearly with input size, and no superlinear growth is observed.

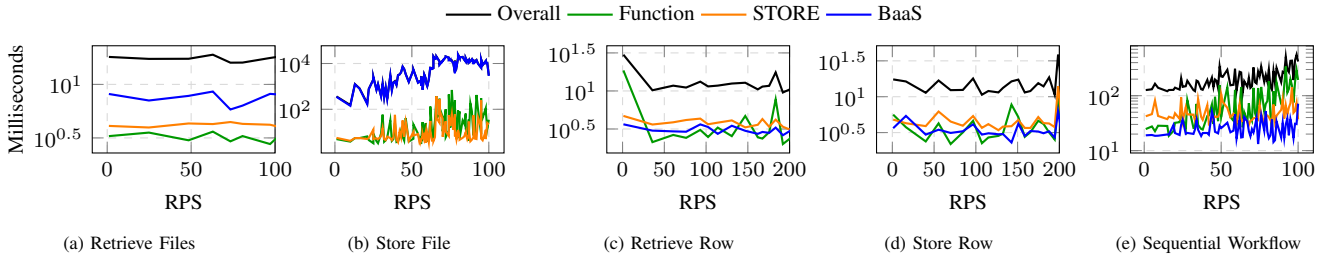
\begin{figure*}[t]
    \begin{subfigure}{0.19\linewidth}
      \begin{tikzpicture}
        \begin{axis}[
          xlabel={RPS}, 
          ylabel={Milliseconds},
          ylabel style={yshift=-3pt,font=\footnotesize},
          xticklabel style={font=\footnotesize},
          xlabel style={font=\footnotesize},
          yticklabel style={font=\footnotesize},
          width=4cm, height=3cm,
          grid=major, grid style={dashed,gray!30},
          mark options={solid},
          legend style={at={(3.25,1.1)},anchor=south,draw=none,legend columns=-1, font=\footnotesize},
          ymode=log,
          xmax=100,
        ]
    
          \addplot+[no marks, color=black, thick] table[x=X, y=overall, col sep=comma] {experiments/scalability/data/read-file.csv};
          \addplot+[no marks, color=green!60!black, thick] table[x=X, y={retrieve-function}, col sep=comma] {experiments/scalability/data/read-file.csv};
          \addplot+[no marks, color=orange, thick] table[x=X, y={store-service},    col sep=comma] {experiments/scalability/data/read-file.csv};
          \addplot+[no marks, color=blue, thick] table[x=X, y={minio.download},   col sep=comma] {experiments/scalability/data/read-file.csv};
          \legend{Overall, Function, STORE, BaaS}
        \end{axis}
      \end{tikzpicture}
      \caption{Retrieve Files}
      \label{fig:workflow_latency}
    \end{subfigure}
       \begin{subfigure}{0.19\linewidth}
        \begin{tikzpicture}
            \begin{axis}[               
                xlabel={RPS},
                ylabel style={yshift=-3pt,font=\footnotesize},
                ylabel={},
                xticklabel style={font=\footnotesize},  
                xlabel style={font=\footnotesize},
                yticklabel style={font=\footnotesize}, 
                width=4cm,                     
                height=3cm,
                grid=major,
                ymode=log,
                scaled y ticks=false,
                grid style={dashed,gray!30},
                mark options={solid}
            ]
           
            \addplot+[no marks, color=black, thick] table[x=X, y=overall, col sep=comma] {experiments/scalability/data/store-file2.csv};
              \addplot+[no marks, color=green!60!black, thick] table[x=X, y={store-function}, col sep=comma] {experiments/scalability/data/store-file2.csv};
              \addplot+[no marks, color=orange, thick] table[x=X, y={store-service},    col sep=comma] {experiments/scalability/data/store-file2.csv};
              \addplot+[no marks, color=blue, thick] table[x=X, y={minio.upload},   col sep=comma] {experiments/scalability/data/store-file2.csv};
            
            \end{axis}
        \end{tikzpicture}
        \caption{Store File}
        \label{fig:workflow_read}
    \end{subfigure}
    \begin{subfigure}{0.19\linewidth}
        \begin{tikzpicture}
            \begin{axis}[               
                xlabel={RPS},
                ylabel style={yshift=-3pt,font=\footnotesize},
                xticklabel style={font=\footnotesize},  
                xlabel style={font=\footnotesize},
                yticklabel style={font=\footnotesize}, 
                ylabel={},
                width=4cm,                    
                height=3cm,
                xmax=200,
                grid=major,
                ymode=log,
                grid style={dashed,gray!30},
                mark options={solid}
            ]
            \addplot+[no marks, color=black, thick]
          table[x=X, y=overall, col sep=comma]
          {experiments/scalability/data/read_row.csv};
          \addplot+[no marks, color=green!60!black, thick] table[x=X, y={retrieve-row-function}, col sep=comma] {experiments/scalability/data/read_row.csv};
          \addplot+[no marks, color=orange, thick] table[x=X, y={store-service},    col sep=comma] {experiments/scalability/data/read_row.csv};
          \addplot+[no marks, color=blue, thick] table[x=X, y={cassandra.get},   col sep=comma] {experiments/scalability/data/read_row.csv};
            \end{axis}
        \end{tikzpicture}
        \caption{Retrieve Row}
        \label{fig:workflow_write}
    \end{subfigure}
    \begin{subfigure}{0.19\linewidth}
        \begin{tikzpicture}
            \begin{axis}[               
                xlabel={RPS},
                ylabel style={yshift=-6pt,font=\footnotesize},
                ylabel={},
                xticklabel style={font=\footnotesize},  
                xlabel style={font=\footnotesize},
                yticklabel style={font=\footnotesize}, 
                width=4cm,                      
                height=3cm,
                grid=major,
                xmax=200,
                ymode=log,
                grid style={dashed,gray!30},
                mark options={solid}
            ]
             \addplot+[no marks, color=black, thick]
          table[x=X, y=overall, col sep=comma]
          {experiments/scalability/data/store-row.csv};
          \addplot+[no marks, color=green!60!black, thick] table[x=X, y={store-row-function}, col sep=comma] {experiments/scalability/data/store-row.csv};
          \addplot+[no marks, color=orange, thick] table[x=X, y={store-service},    col sep=comma] {experiments/scalability/data/store-row.csv};
          \addplot+[no marks, color=blue, thick] table[x=X, y={cassandra.save},   col sep=comma] {experiments/scalability/data/store-row.csv};
            \end{axis}
        \end{tikzpicture}
        \caption{Store Row}
        \label{fig:workflow_throughput}
    \end{subfigure}
    \begin{subfigure}{0.19\linewidth}
        \begin{tikzpicture}
            \begin{axis}[               
                xlabel={RPS},
                ylabel style={yshift=-6pt,font=\footnotesize},
                ylabel={},
                xticklabel style={font=\footnotesize},  
                xlabel style={font=\footnotesize},
                yticklabel style={font=\footnotesize}, 
                width=4cm,                      
                height=3cm,
                grid=major,
                ymode=log,
                grid style={dashed,gray!30},
                mark options={solid}
            ]
             \addplot+[no marks, color=black, thick]
          table[x=X, y=overall, col sep=comma]
          {experiments/scalability/data/workflow.csv};
          \addplot+[no marks, color=green!60!black, thick] table[x=X, y={functions}, col sep=comma] {experiments/scalability/data/workflow.csv};
          \addplot+[no marks, color=orange, thick] table[x=X, y={store-service},    col sep=comma] {experiments/scalability/data/workflow.csv};
          \addplot+[no marks, color=blue, thick] table[x=X, y={storage},   col sep=comma] {experiments/scalability/data/workflow.csv};
            \end{axis}
        \end{tikzpicture}
        \caption{Sequential Workflow}
        \label{fig:workflow_throughput}
    \end{subfigure}
    \caption{Scalability of serverless workflows with varying request per seconds (RPS): }
    \label{fig:scalability}
\end{figure*}
\begin{figure}[t]
    \begin{subfigure}{0.49\linewidth}
        \begin{tikzpicture}
            \begin{groupplot}[
              group style={group size=1 by 2, vertical sep=12pt},
              width=4.5cm, height=2.4cm,
              axis lines=box, ytick pos=left,
              yticklabel style={font=\footnotesize},
              xticklabel style={font=\footnotesize},  
              grid=major, grid style={dashed,gray!30},
              ylabel={CPU Usage (millicores)},
            ]
            
            \nextgroupplot[
              ylabel style={font=\footnotesize, xshift=-22pt},
              legend style={at={(1.15,2.15)}, anchor=north, draw=none, font=\footnotesize, legend columns=3},
              ytick distance=1500,
              xmax=100,
            ]
            \addplot+[no marks, color=black, thick]
              table[x=X, y={store}, col sep=comma]{experiments/scalability/res_usage/data/read_file_store_cpu.csv};
            \addlegendentry{Retrieve File}
            
            \addplot+[no marks, color=blue, thick]
              table[x=X, y={store}, col sep=comma]{experiments/scalability/res_usage/data/store_file2_store_cpu.csv};
            \addlegendentry{Store File}

            \addplot+[no marks, color=red, thick]
              table[x=X, y={store}, col sep=comma]{experiments/scalability/res_usage/data/workflow_store_cpu.csv};
            \addlegendentry{Sequential Workflow}

            \addlegendimage{no markers, draw=orange, thick}
            \addlegendentry{Retrieve Row}

            \addlegendimage{no markers, draw=green!60!black, thick}
            \addlegendentry{Store Row}
            
            \nextgroupplot[axis lines=box, ytick pos=left, 
            xmax=200,
            ytick distance=400,
            ylabel={},
            xlabel={RPS},
            xlabel style={font=\scriptsize}]
            \addplot+[no marks, color=orange, thick]
              table[x=X, y={store}, col sep=comma]{experiments/scalability/res_usage/data/read_row_store_cpu.csv};
            
            \addplot+[no marks, color=green!60!black, thick]
              table[x=X, y={store}, col sep=comma]{experiments/scalability/res_usage/data/store_row_store_cpu.csv};

            
            \end{groupplot}
            
            \end{tikzpicture}
        \caption{CPU}
    \end{subfigure}
    \hfill
    \begin{subfigure}{0.49\linewidth}
        \begin{tikzpicture}
            \begin{groupplot}[
              group style={group size=1 by 2, vertical sep=12pt},
              width=4.5cm, height=2.4cm,
              axis lines=box, ytick pos=left,
              yticklabel style={font=\footnotesize},
              xticklabel style={font=\footnotesize},  
              grid=major, grid style={dashed,gray!30},
              ylabel={Memory (MB)}
            ]
            
            \nextgroupplot[
              xmax=100,
              ytick distance=400,
              ylabel style={font=\footnotesize, xshift=-20pt},
            ]
            \addplot+[no marks, color=black, thick]
              table[x=X, y={store}, col sep=comma]{experiments/scalability/res_usage/data/read_file_store_mem.csv};
            
            \addplot+[no marks, color=blue, thick]
              table[x=X, y={store}, col sep=comma]{experiments/scalability/res_usage/data/store_file2_store_mem.csv};
            
            \nextgroupplot[axis lines=box, ytick pos=left, ylabel={},
            xmax=200,
            xlabel={RPS},
            xlabel style={font=\scriptsize}
            ]
            \addplot+[no marks, color=orange, thick]
              table[x=X, y={store}, col sep=comma]{experiments/scalability/res_usage/data/read_row_store_mem.csv};

            \addplot+[no marks, color=green!60!black, thick]
              table[x=X, y={store}, col sep=comma]{experiments/scalability/res_usage/data/store_row_store_mem.csv};
            
            \end{groupplot}
            
            \end{tikzpicture}
        \caption{RAM}
    \end{subfigure}
    \caption{Resource usage of Self Provisioning Service where: (a) CPU, and (b) RAM. Solid line is Self Provisioning Service}
    \label{fig:scalability_resusage}
\end{figure}
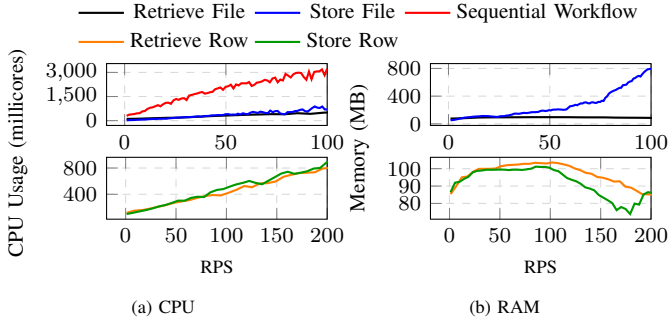

\subsubsection{Resource Usage}

\cref{fig:performance-resources} shows that CPU and memory consumption remain well within the defined success criteria, with no evidence of superlinear growth.
\textsc{STORE} CPU usage is consistently low and stable, while function CPU usage increases moderately with request size but stays in the range of $400$–$600$~millicores for most workloads. Retrieving files is the most demanding case, peaking at $1609$~millicores for the service. Memory consumption is minimal after startup ($36$MB) and scales up to $640$~MB in the most demanding case (store file). These fluctuations are primarily attributed to garbage-collection cycles, and overall, both CPU and memory trends confirm efficient resource behavior across file- and table-based workloads.

\subsection{Scalability Results} 
\subsubsection{Latency} \cref{fig:scalability} shows that \textsc{STORE} overhead remains between 3 and 514~ms across all scalability experiments

\emph{File-based workloads:} For storing files, BaaS (MinIO) dominates execution time, while the service and function contribute only a small share. Absolute overhead fluctuates but remains low (mean $71$~ms), and relative overhead decreases with scale, averaging $1.9\%$ and peaking at $15.6\%$. Retrieving files shows even lower absolute overhead (mean $4.3$~ms) with stable scaling and a relative overhead that decreases from $19\%$ toward negligible levels.  

\emph{Tabular workloads:} For storing rows, execution is balanced across service, function, and BaaS (Cassandra), with small absolute overhead (mean $4.6$~ms) and stable relative overhead ($29\%$). Retrieving rows shows similar trends (mean $4.2$~ms), though variability is higher; relative overhead again decreases with scale.  

\emph{Workflow workloads:} The full workflow introduces additional orchestration overhead, but absolute values remain modest (mean $101$~ms). Overhead is higher (up to $38\%$) due to small request sizes, yet consistently decreases with scale.

\subsubsection{Resource Usage} \cref{fig:scalability_resusage} shows that the overall, CPU usage satisfies the linearity criterion across all scenarios, while memory usage is generally stable and predictable.

\emph{CPU usage:} Across all experiments, CPU consumption scales linearly with request rate for both the service and function components.  Experiments show that CPU demand grows proportionally with load and never exceeds the ``at most linear''. The workflow scenario exhibits the highest demand ($3306$~millicores for the service, $1699$~millicores for functions), while file storage shows the largest service/function imbalance ($1708$ vs.\ $467$~millicores).

\emph{Memory usage:} Function memory remains stable across all workloads, while service memory is dominated by request-buffering in file storage. 
Most use cases show only an initial allocation bump from the BaaS (Quarkus Native) ($42$~MB) followed by stable scaling. The steepest increase occurs with file storage, where per-request buffers raise service memory to a maximum of $850$~MB. This could be further reduced by introducing buffer-pooling, but it never exceeds the ``at most linear''.

\subsection{Line of Code Results}

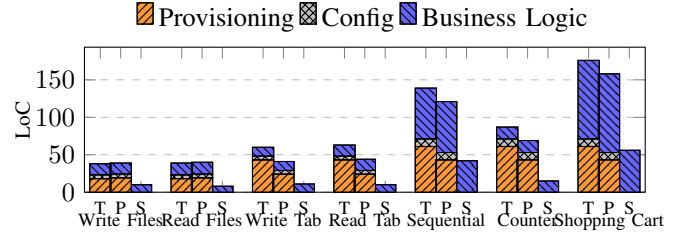
\begin{figure}
    \centering
    \begin{tikzpicture}
    \begin{axis}[
        ybar stacked,
        ymin=0,
        bar width=8pt,
        ylabel={LoC},
        ylabel style={yshift=-5pt,font=\footnotesize},
        ymajorgrids=true,
        grid style=dashed,
        height=3.5cm,
        width=9cm,
        xtick=data,
        symbolic x coords={
          SF-Terraform,SF-Pulumi,SF-STORE,,
          RF-Terraform,RF-Pulumi,RF-STORE,,
          STD-Terraform,STD-Pulumi,STD-STORE,,
          RTD-Terraform,RTD-Pulumi,RTD-STORE,,
          SWF-Terraform,SWF-Pulumi,SWF-STORE,,
          CWF-Terraform,CWF-Pulumi,CWF-STORE,,
          SCW-Terraform,SCW-Pulumi,SCW-STORE
        },
        xticklabels={
          T, P, S, 
          T, P, S, 
          T, P, S, 
          T, P, S, 
          T, P, S, 
          T, P, S, 
          T, P, S, 
        },
        xticklabel style={font=\scriptsize},
        enlarge x limits=0.03,
        legend style={at={(0.5,1.35)}, anchor=north, draw=none, legend columns=-1},
        clip=false,
    ]
        \addplot [fill=orange!80, postaction={pattern=north east lines}]
            table [x=Label, y=Provisioning] {\iacost};
    
        \addplot [fill=gray!50, postaction={pattern=crosshatch}]
            table [x=Label, y=Config] {\iacost};
    
        \addplot [fill=blue!60, postaction={pattern=north west lines}]
            table [x=Label, y=Business] {\iacost};
    
        \legend{Provisioning, Config, Business Logic}
        
        \node at (axis cs:{SF-Pulumi}, -15) [anchor=north, font=\scriptsize] {Write Files};
        \node at (axis cs:{RF-Pulumi}, -15) [anchor=north, font=\scriptsize] {Read Files};
        
        \node at (axis cs:{STD-Pulumi}, -15) [anchor=north, font=\scriptsize] {Write Tab};
        \node at (axis cs:{RTD-Pulumi}, -15) [anchor=north, font=\scriptsize] {Read Tab};
        
        \node at (axis cs:{SWF-Pulumi}, -15) [anchor=north, font=\scriptsize] {Sequential};
        \node at (axis cs:{CWF-Pulumi}, -15) [anchor=north, font=\scriptsize] {Counter};
        \node at (axis cs:{SCW-Pulumi}, -15) [anchor=north, font=\scriptsize] {Shopping Cart};
    \end{axis}
    \end{tikzpicture}

    \caption{Line of Codes for (T)erraform, (P)ulumi and (S)TORE}
    \label{fig:loc}
\end{figure}

We compare developer effort by summing the lines of code required for BaaS provisioning, connection configuration, and business logic for each tested use case.
STORE requires less implementation effort from 64.6\% to 84.1\% compared to Terraform and Pulumi. 
In addition to the designed workflows, we add the following real-world use cases: \emph{Counter Workflow} employs the \texttt{map} interface to manage a quota counter with optimistic concurrency control across three steps. \emph{Shopping Cart Workflow} combines validation, cart update, analytics, and enrichment, and integrates both file/tabular storage with the \texttt{map} interface.
As shown in \cref{fig:loc}, our STORE prototype completely eliminates configuration and provisioning code and consistently requires fewer lines of code across all use cases, ranging from only eight lines (\textit{Retrieving Files}) to 56 lines (\textit{Shopping Cart Workflow}). In contrast, Terraform requires 38–176 lines of code, while Pulumi requires 39–158 lines. The smallest gap occurs in \textit{Storing Files}, where STORE uses 28 fewer lines than Terraform (10 vs. 38). The largest gap appears in the \textit{Shopping Cart Workflow}, where STORE reduces the effort by 120 lines (56 vs. 176), cutting implementation effort to less than one third of IaC-based approaches.  

STORE also achieves notable reductions when considering only function bodies, with absolute savings from 1 line (\textit{Storing Tabular Data} and \textit{Counter Workflow}) up to 49 lines (\textit{Shopping Cart Workflow}). Improvements span from 6.2\% to 50.0\%. Overall, the results confirm that STORE enables significantly more concise implementations than well-established commercial IaC platforms such as Terraform and Pulumi.
We acknowledge that lines of code are only a coarse proxy for developer effort and do not capture all aspects of complexity or maintainability.

\pgfplotstableread[col sep=comma]{
UseCase,Label,Provisioning,Config,Business
Storing Files,SF-Terraform,18,5,15
Storing Files,SF-Pulumi,19,5,15
Storing Files,SF-STORE,0,0,10
Retrieving Files,RF-Terraform,18,5,16
Retrieving Files,RF-Pulumi,19,5,16
Retrieving Files,RF-STORE,0,0,8
Storing Tabular Data,STD-Terraform,43,5,12
Storing Tabular Data,STD-Pulumi,24,5,12
Storing Tabular Data,STD-STORE,0,0,11
Retrieving Tabular Data,RTD-Terraform,43,5,15
Retrieving Tabular Data,RTD-Pulumi,24,5,15
Retrieving Tabular Data,RTD-STORE,0,0,10
Sequential Workflow,SWF-Terraform,61,10,68
Sequential Workflow,SWF-Pulumi,43,10,68
Sequential Workflow,SWF-STORE,0,0,42
Counter Workflow,CWF-Terraform,61,10,16
Counter Workflow,CWF-Pulumi,43,10,16
Counter Workflow,CWF-STORE,0,0,15
Shopping Cart Workflow,SCW-Terraform,61,10,105
Shopping Cart Workflow,SCW-Pulumi,43,10,105
Shopping Cart Workflow,SCW-STORE,0,0,56
}\iacost

\section{Conclusion}\label{conclusion}

In this paper, we present STORE, a novel architecture that enables Self-Provisioning storage, alleviating developers of the burden of manual storage selection, provisioning, and connection configuration. Moreover, STORE provides dynamic storage selection based on data shape and access patterns, as well as zero-touch configuration, enabling self-provisioned storage without manual intervention. We evaluated a STORE implementation against state-of-the-art IaC frameworks and demonstrated that it reduces development effort by up to 84\% for a wide range of serverless use cases, while maintaining performance and scalability comparable to directly employing underlying storage solutions with relative overheads as low as 0.19\% and a low resource profile.

In the future, we plan to extend dynamic storage selection to specify a maximum number of storage locations to retain simultaneously, to reduce switching costs, and a holdout duration after which unused locations are cleared. Additionally, we aim to extend STORE to operate across the edge-cloud continuum, potentially introducing a dimension of storage locality. We also plan to extend the solution to support multi-cloud environments, enabling the provisioning of storage solutions across multiple cloud providers. 

\bibliographystyle{IEEEtran}
\bibliography{references}

\begin{thebibliography}{10}
\providecommand{\url}[1]{#1}
\csname url@samestyle\endcsname
\providecommand{\newblock}{\relax}
\providecommand{\bibinfo}[2]{#2}
\providecommand{\BIBentrySTDinterwordspacing}{\spaceskip=0pt\relax}
\providecommand{\BIBentryALTinterwordstretchfactor}{4}
\providecommand{\BIBentryALTinterwordspacing}{\spaceskip=\fontdimen2\font plus
\BIBentryALTinterwordstretchfactor\fontdimen3\font minus \fontdimen4\font\relax}
\providecommand{\BIBforeignlanguage}[2]{{%
\expandafter\ifx\csname l@#1\endcsname\relax
\typeout{** WARNING: IEEEtran.bst: No hyphenation pattern has been}%
\typeout{** loaded for the language `#1'. Using the pattern for}%
\typeout{** the default language instead.}%
\else
\language=\csname l@#1\endcsname
\fi
#2}}
\providecommand{\BIBdecl}{\relax}
\BIBdecl

\bibitem{scf}
S.~Nastic, P.~Raith, A.~Furutanpey, T.~Pusztai, and S.~Dustdar, ``A serverless computing fabric for edge \& cloud,'' in \emph{2022 IEEE 4th International Conference on Cognitive Machine Intelligence (CogMI)}, 2022, pp. 1--12.

\bibitem{CloudProgrammingSimplified}
E.~Jonas, J.~Schleier-Smith, V.~Sreekanti, C.-C. Tsai, A.~Khandelwal, Q.~Pu, V.~Shankar, J.~Carreira, K.~Krauth, N.~Yadwadkar, J.~E. Gonzalez, R.~A. Popa, I.~Stoica, and D.~A. Patterson, ``Cloud programming simplified: A berkeley view on serverless computing,'' 2019.

\bibitem{goldfish2024}
C.~Marcelino, J.~Shahhoud, and S.~Nastic, ``Goldfish: Serverless actors with short-term memory state for the edge-cloud continuum,'' in \emph{Proceedings of the 14th International Conference on the Internet of Things}, ser. IoT '24.\hskip 1em plus 0.5em minus 0.4em\relax New York, NY, USA: ACM, 2024.

\bibitem{PerformanceIsolation}
R.~Gajanin, C.~Marcelino, and S.~Nastic, ``Performance isolation for serverless functions,'' \emph{IEEE Transactions on Services Computing}, vol.~18, no.~6, pp. 4408--4424, 2025.

\bibitem{Lumos2025}
C.~Marcelino, N.~Krennmair, T.~W. Pusztai, and S.~Nastic, ``Lumos: Performance characterization of webassembly as a serverless runtime in the edge-cloud continuum,'' in \emph{Proceedings of the 15th International Conference on the Internet of Things}, ser. IOT '25, 2025, p. 113–121.

\bibitem{Databelt2025}
C.~Marcelino, L.~Guelmino, T.~Pusztai, and S.~Nastic, ``Databelt: A continuous data path for serverless workflows in the 3d compute continuum,'' \emph{Journal of Systems Architecture}, vol. 168, p. 103577, 2025.

\bibitem{ristov2024code}
S.~Ristov, S.~Brandacher, M.~Hautz, M.~Felderer, and R.~Breu, ``Code: code once, deploy everywhere serverless functions in federated faas,'' \emph{Future Generation Computer Systems}, vol. 160, pp. 442--456, 2024.

\bibitem{sokolowski2021automating}
D.~Sokolowski, P.~Weisenburger, and G.~Salvaneschi, ``Automating serverless deployments for devops organizations,'' in \emph{Proceedings of the 29th ACM Joint Meeting on European Software Engineering Conference and Symposium on the Foundations of Software Engineering}, ser. ESEC/FSE 2021, 2021, p. 57–69.

\bibitem{2024selfprovisioningInfrastructure}
S.~Nastic, ``Self-provisioning infrastructures for the next generation serverless computing,'' \emph{SN Computer Science}, vol.~5, no.~6, pp. 678 -- 693, 2024.

\bibitem{terraformTerraformHashiCorp}
``{T}erraform by {H}ashi{C}orp --- terraform.io,'' \url{https://www.terraform.io/}, [Accessed 27-03-2025].

\bibitem{pulumiPulumiInfrastructure}
``{P}ulumi - {I}nfrastructure as {C}ode, {S}ecrets {M}anagement, and {A}{I} --- pulumi.com,'' \url{https://www.pulumi.com/}, [Accessed 01-04-2025].

\bibitem{awsCloudFormation}
\BIBentryALTinterwordspacing
{Amazon Web Services, Inc.}, ``Aws cloudformation,'' 2025, accessed: 2025-09-05. [Online]. Available: \url{https://aws.amazon.com/cloudformation/}
\BIBentrySTDinterwordspacing

\bibitem{YussupovBKL20}
\BIBentryALTinterwordspacing
V.~Yussupov, U.~Breitenb{\"{u}}cher, A.~Kaplan, and F.~Leymann, ``{SEAPORT:} assessing the portability of serverless applications,'' in \emph{Proceedings of the 10th International Conference on Cloud Computing and Services Science, {CLOSER} 2020, Prague, Czech Republic, May 7-9, 2020}, D.~Ferguson, M.~Helfert, and C.~Pahl, Eds.\hskip 1em plus 0.5em minus 0.4em\relax {SCITEPRESS}, 2020, pp. 456--467. [Online]. Available: \url{https://doi.org/10.5220/0009574104560467}
\BIBentrySTDinterwordspacing

\bibitem{WursterBFKLSS20}
\BIBentryALTinterwordspacing
M.~Wurster, U.~Breitenb{\"{u}}cher, M.~Falkenthal, C.~Krieger, F.~Leymann, K.~Saatkamp, and J.~Soldani, ``The essential deployment metamodel: a systematic review of deployment automation technologies,'' \emph{{SICS} Softw.-Intensive Cyber Phys. Syst.}, vol.~35, no. 1-2, pp. 63--75, 2020. [Online]. Available: \url{https://doi.org/10.1007/s00450-019-00412-x}
\BIBentrySTDinterwordspacing

\bibitem{DBLP:conf/icde/KesavanGTSM23}
R.~Kesavan, D.~Gay, D.~Thevessen, J.~Shah, and C.~Mohan, ``Firestore: The nosql serverless database for the application developer,'' in \emph{39th {IEEE} International Conference on Data Engineering, {ICDE} 2023, Anaheim, CA, USA, April 3-7, 2023}.\hskip 1em plus 0.5em minus 0.4em\relax {IEEE}, 2023, pp. 3376--3388.

\bibitem{DBLP:conf/sigmod/SwensonKPTLSBBT25}
J.~Swenson, A.~Kimball, R.~k. Poss, R.~Taft, J.~Lim, A.~Storm, S.~Bhola, P.~Bulkley-Logston, P.~Tatlow, R.~Harding, R.~Shamim, A.~Maru, and I.~Sharif, ``Cockroachdb serverless: Sub-second scaling from zero with multi-region cluster virtualization,'' in \emph{Companion of the 2025 International Conference on Management of Data}, ser. SIGMOD/PODS '25, 2025, p. 648–661.

\bibitem{amazonCloudDevelopment}
``{C}loud {D}evelopment {F}ramework - {A}{W}{S} {C}loud {D}evelopment {K}it - {A}{W}{S} --- aws.amazon.com,'' \url{https://aws.amazon.com/cdk/}, [Accessed 14-09-2025].

\bibitem{Allam_2025}
\BIBentryALTinterwordspacing
H.~Allam, ``Intent-based infrastructure: Moving beyondiac to self-describing systems,'' \emph{International Journal of Artificial Intelligence, Data Science, and Machine Learning}, vol.~6, no.~1, p. 124–136, Jan. 2025. [Online]. Available: \url{https://ijaidsml.org/index.php/ijaidsml/article/view/182}
\BIBentrySTDinterwordspacing

\bibitem{DBLP:conf/nof/BezahafHBDBKH19}
M.~Bezahaf, M.~P. Hernandez, L.~Bardwell, E.~Davies, M.~Broadbent, D.~King, and D.~Hutchison, ``Self-generated intent-based system,'' in \emph{2019 10th International Conference on Networks of the Future (NoF)}, 2019, pp. 138--140.

\bibitem{DBLP:journals/sensors/AndradeHozWA24}
\BIBentryALTinterwordspacing
J.~Andrade{-}Hoz, Q.~Wang, and J.~M. Alcaraz{-}Calero, ``Infrastructure-wide and intent-based networking dataset for 5g-and-beyond ai-driven autonomous networks,'' \emph{Sensors}, vol.~24, no.~3, p. 783, 2024. [Online]. Available: \url{https://doi.org/10.3390/s24030783}
\BIBentrySTDinterwordspacing

\bibitem{DBLP:conf/ucc/BhattacharjeeBG18}
A.~Bhattacharjee, Y.~Barve, A.~Gokhale, and T.~Kuroda, ``A model-driven approach to automate the deployment and management of cloud services,'' in \emph{2018 IEEE/ACM International Conference on Utility and Cloud Computing Companion (UCC Companion)}, 2018, pp. 109--114.

\bibitem{DBLP:conf/models/SandobalinIA19}
J.~Sandobalin, E.~Insfran, and S.~Abrahão, ``Argon: A model-driven infrastructure provisioning tool,'' in \emph{2019 ACM/IEEE 22nd International Conference on Model Driven Engineering Languages and Systems Companion (MODELS-C)}, 2019, pp. 738--742.

\bibitem{darklangDarklang}
``{D}arklang --- darklang.com,'' \url{https://darklang.com/}, [Accessed 27-03-2025].

\bibitem{winglangWingProgramming}
``{W}ing {P}rogramming {L}anguage for the cloud | {W}ing --- winglang.io,'' \url{https://www.winglang.io/}, [Accessed 27-03-2025].

\bibitem{DBLP:conf/seke/QianZ20}
C.~Qian and W.~Zhu, ``{F(X)-MAN:} an algebraic and hierarchical composition model for function-as-a-service,'' in \emph{The 32nd International Conference on Software Engineering and Knowledge Engineering, {SEKE} 2020, {KSIR} Virtual Conference Center, USA, July 9-19, 2020}, R.~Garc{\'{\i}}a{-}Castro, Ed.\hskip 1em plus 0.5em minus 0.4em\relax {KSI} Research Inc., 2020, pp. 210--215.

\bibitem{DBLP:conf/cloudcom/MoCCL23}
D.~Mo, R.~Cordingly, D.~Chinn, and W.~Lloyd, ``Addressing serverless computing vendor lock-in through cloud service abstraction,'' in \emph{2023 IEEE International Conference on Cloud Computing Technology and Science (CloudCom)}, 2023, pp. 193--199.

\bibitem{apacheApacheLibcloud}
T.~A.~S. Foundation, ``{A}pache {L}ibcloud is a standard {P}ython library that abstracts away differences among multiple cloud provider {A}{P}{I}s --- libcloud.apache.org,'' \url{https://libcloud.apache.org/}, [Accessed 15-09-2025].

\bibitem{apacheApacheJcloudsxAE}
``{A}pache jclouds :: {H}ome --- jclouds.apache.org,'' \url{https://jclouds.apache.org/}, [Accessed 15-09-2025].

\bibitem{DBLP:conf/icsca/SameaAAKR19}
F.~Samea, F.~Azam, M.~W. Anwar, M.~Khan, and M.~Rashid, ``A uml profile for multi-cloud service configuration (umlpmsc) in event-driven serverless applications,'' in \emph{Proceedings of the 2019 8th International Conference on Software and Computer Applications}, ser. ICSCA '19, 2019, p. 431–435.

\bibitem{LarcherGNR24}
\BIBentryALTinterwordspacing
T.~Larcher, P.~Gritsch, S.~Nastic, and S.~Ristov, ``Baasless: Backend-as-a-service (baas)-enabled workflows in federated serverless infrastructures,'' \emph{{IEEE} Trans. Cloud Comput.}, vol.~12, no.~4, pp. 1088--1102, 2024. [Online]. Available: \url{https://doi.org/10.1109/TCC.2024.3439268}
\BIBentrySTDinterwordspacing

\bibitem{DBLP:conf/icsoc/RistovHGNPF24}
S.~Ristov, M.~Hautz, P.~Gritsch, S.~Nastic, R.~Prodan, and M.~Felderer, ``Storeless: Serverless workflow scheduling with federated storage in sky computing,'' in \emph{Service-Oriented Computing}, W.~Gaaloul, M.~Sheng, Q.~Yu, and S.~Yangui, Eds.\hskip 1em plus 0.5em minus 0.4em\relax Singapore: Springer Nature Singapore, 2025, pp. 35--44.

\bibitem{280754}
\BIBentryALTinterwordspacing
M.~Elhemali, N.~Gallagher, N.~Gordon, J.~Idziorek, R.~Krog, C.~Lazier, E.~Mo, A.~Mritunjai, S.~Perianayagam, T.~Rath, S.~Sivasubramanian, J.~C.~S. III, S.~Sosothikul, D.~Terry, and A.~Vig, ``Amazon {DynamoDB}: A scalable, predictably performant, and fully managed {NoSQL} database service,'' in \emph{2022 USENIX Annual Technical Conference (USENIX ATC 22)}.\hskip 1em plus 0.5em minus 0.4em\relax Carlsbad, CA: USENIX Association, Jul. 2022, pp. 1037--1048. [Online]. Available: \url{https://www.usenix.org/conference/atc22/presentation/elhemali}
\BIBentrySTDinterwordspacing

\bibitem{DBLP:conf/sigmod/ThomsonDWRSA12}
\BIBentryALTinterwordspacing
A.~Thomson, T.~Diamond, S.-C. Weng, K.~Ren, P.~Shao, and D.~J. Abadi, ``Calvin: fast distributed transactions for partitioned database systems,'' in \emph{Proceedings of the 2012 ACM SIGMOD International Conference on Management of Data}, ser. SIGMOD '12.\hskip 1em plus 0.5em minus 0.4em\relax New York, NY, USA: Association for Computing Machinery, 2012, p. 1–12. [Online]. Available: \url{https://doi.org/10.1145/2213836.2213838}
\BIBentrySTDinterwordspacing

\bibitem{minio}
{MinIO, Inc.}, ``Minio: High-performance, s3-compatible, kubernetes-native object storage,'' \url{https://www.min.io/}, MinIO, Inc., 2025, accessed: 2025-09-05.

\bibitem{apacheCassandra}
{The Apache Software Foundation}, ``Apache cassandra: Open-source nosql distributed database,'' \url{https://cassandra.apache.org/}, Apache Software Foundation, 2025, accessed: 2025-09-05.

\bibitem{redis}
{Redis Ltd.}, ``Redis: Open-source, in-memory data store and cache,'' \url{https://redis.io/}, Redis Ltd., 2025, accessed: 2025-09-05.

\bibitem{jaeger}
{The Jaeger Contributors and Community}, ``Jaeger: Open-source, distributed tracing platform,'' \url{https://www.jaegertracing.io/}, Jaeger Community (CNCF Project), 2025, accessed: 2025-09-05.

\bibitem{prometheus}
{The Prometheus Authors}, ``Prometheus: Open-source monitoring and alerting toolkit,'' \url{https://prometheus.io/}, Cloud Native Computing Foundation, 2025, accessed: 2025-09-05.

\bibitem{grafana}
{Grafana Labs}, ``Grafana: Open-source analytics and monitoring platform,'' \url{https://grafana.com/}, Grafana Labs, 2025, accessed: 2025-09-05.

\end{thebibliography}

\end{document}